\documentclass{PoS}

\newcommand{\be}{\begin{equation}}
\newcommand{\ee}{\end{equation}}
\newcommand{\bea}{\begin{eqnarray}}
\newcommand{\eea}{\end{eqnarray}}
\newcommand{\nn}{\nonumber}

\newcommand{\one}{\mbox{\bf 1}}
\def\Tr{{\mbox{Tr}}}
\def\hx{\hat{x}}
\def\hy{\hat{y}}
\def\hm{\hat{m}}
\def\hmu{\hat{\mu}}

\title{Determination of $F_{\pi}$ from Distributions of Dirac Operator
  Eigenvalues with Imaginary Density} 

\ShortTitle{$F_\pi$ from Dirac eigenvalues at imaginary density}

\author{\speaker{G.~Akemann}
\\
        Department of Mathematical Sciences \& BURSt Research Centre\\
Brunel University West London, 
Uxbridge UB8 3PH, United Kingdom\\
and\\
Isaac Newton Institute for Mathematical Sciences\\
20 Clarkson Road, Cambridge CB3 0EH, United Kingdom\\
        E-mail: \email{gernot.akemann@brunel.ac.uk}
}

\author{P.~H.~Damgaard\\
The Niels Bohr Institute \& The Niels Bohr International Academy\\
Blegdamsvej 17, DK-2100 Copenhagen, Denmark
\\
        E-mail: \email{phdamg@nbi.dk}}

\abstract{In the $\epsilon$-regime of lattice QCD one can get an accurate
measurement 
of the pion decay constant $F_{\pi}$ by monitoring how just one single Dirac
operator eigenvalue splits into two when subjected to two different 
external vector sources. 
Because we choose imaginary chemical potentials our Dirac eigenvalues remain
real. Based on the relevant chiral Random Two-Matrix Theory
we derive individual eigenvalue distributions in terms of density correlations
functions to leading order in the finite-volume
$\epsilon$-expansion. As a simple byproduct we also show how the
associated individual Dirac eigenvalue distributions and their correlations
can be computed directly from the effective chiral Lagrangian. 
}

\FullConference{The XXV International Symposium on Lattice Field Theory\\
                 July 30-4 August 2007\\
                 Regensburg, Germany}

\begin{document}
\section{Introduction}

The distribution of individual Dirac operator eigenvalues 
 have become a popular
tool since their calculation \cite{DNW} 
in the chiral Random Matrix Theory that is equivalent to the 
leading-order expression for the QCD partition function in
the $\epsilon$-regime \cite{SV93}. The
analytical expressions distinguish very clearly between different gauge
theories and different sectors of topology, as first shown in
\cite{florida}. This has by now been 
verified by many different groups using different versions of fermions with 
different levels of chiral symmetry on the lattice. 
It has also become clear how to derive the same expression in the 
$\epsilon$-regime of chiral perturbations theory ($\epsilon\chi$PT)
\cite{ADp}.  
 
Individual eigenvalue distributions provide perhaps the most efficient tool to
extract 
one of the low energy constants (LEC) in $\chi$PT, the infinite-volume
chiral condensate $\Sigma$. 
Here we extend this analysis to the second LEC in line, the pion decay
constant  
$F_\pi$, exploiting the fact that a nonvanishing chemical potential $\mu$ 
couples to $F_\pi$ to leading order in the $\epsilon$-expansion \cite{TV}. 
This method has been first suggested for imaginary isospin 
chemical potential with two different sets of Dirac eigenvalues \cite{DHSST}. 
The advantage over real $\mu$ \cite{AW} is that the Dirac operator
retains its anti-hermiticity,
allowing for unquenched simulations without
encountering any sign-problems, and with greatly reduced computer
efforts associated with the computation of the lowest eigenvalues.
The proposal \cite{DHSST} was based on the 2-point spectral correlation
function computed from $\epsilon\chi$PT
and verified their prediction on quenched and unquenched Lattice data. 
This was generalised in \cite{ADOS} where all spectral correlations where
computed analytically from the shown equivalence with a
corresponding chiral Random two-Matrix Theory (chR2MT) with $\mu_{1,2}$. 
The advantage here is that partial
quenching is possible, by setting one of the $\mu_j$ to zero. 
Hence existing configurations with $\mu=0$ can be used to measure $F_\pi$.
This idea was most recently applied to unquenched
chiral fermions in \cite{TDG}.
Here we present first results for individual Dirac eigenvalue distributions. 

In section \ref{RMTchPT} we introduce the chR2MT and its
corresponding $\epsilon\chi$PT. Section \ref{results} presents our 
results in a general setting, which is then illustrated pictorially 
in the simplest case, the quenched isospin densities in 
section \ref{ex}. Section \ref{conc} gives
our conclusions and comments on other results.

\section{RMT and $\chi$PT with imaginary chemical potential}\label{RMTchPT}

We start by defining the chR2MT for imaginary chemical
potentials introduced and solved 
in \cite{ADOS}
\bea
{\cal Z}_{chR2MT}
&\sim&
 \int d\Phi  d\Psi~ \exp\left[-{N}{\rm Tr}\left(\Phi^{\dagger}
\Phi + \Psi^{\dagger}\Psi\right)\right]  
\prod_{f=1}^{N_f} \det[{\cal D}(\mu_f) + m_{f}] \ .
\label{ZNf}
\eea
The anti-hermitian Dirac matrix ${\cal D}$ 
is given in terms of two complex, rectangular 
random matrices $\Phi$ and $\Psi$ of size $N\times (N+\nu)$
with Gaussian measure: 
\bea
{\mathcal D}(\mu_f) = \left( \begin{array}{cc}
0 & i \Phi + i \mu_f \Psi \\
i \Phi^{\dagger} + i \mu_f \Psi^{\dagger} & 0
\end{array} \right) ~.
\eea
Here 
$\nu$ corresponds to fixed gauge field topology in the usual way.
In the following we restrict ourselves to the case of only two 
different chemical potentials, 
${\mathcal D}(\mu_{1,2})\equiv {\mathcal  D}_{1,2} $, 
with $N_{1,2}$ flavours each.
Referring to ref. \cite{ADOS} for details, we can write down
the corresponding eigenvalue representation:
\bea
{\cal Z}_{chR2MT}
&=& \int_0^{\infty} \prod_i^N\left(dx_idy_i (x_iy_i)^{\nu+1}
\prod_{f1=1}^{N_1} (x_i^2+m_{f1}^2)
\prod_{f2=1}^{N_2} (y_i^2+m_{f2}^2) \right) \cr
&&\times\ \Delta_N(\{x^2\})\Delta_N(\{y^2\})\det\left[I_{\nu}(2 d N x_i y_j)
\right] 
e^{-N \sum_i c_1 x_i^2 + c_2 y_i^2 } ,
\label{evrep}
\eea
where $x_i$ and $y_i$ are real positive eigenvalues of the matrices
$\Phi+\mu_{1,2}\Psi$, respectively. The constants $c_{1,2}$ and $d$ depend on 
$\mu_{1,2}$ (see \cite{ADOS}). For later convenience we 
abbreviate the integrand or joint probability distribution function (jpdf) by 
${\cal P}(\{x\},\{y\})$.

If we take the large-$N$ limit and identify $Nx_i\to V\Sigma x_i\equiv\hx$,
$Nm_{f1}\to V\Sigma m_{f1}\equiv\hm_{f1}$,
$2N\mu^2_1\to VF_\pi^2\mu^2_1\equiv\hat{\mu}_1^2$, 
and similarly for the second set $N_2$, the
partition function eq. (\ref{evrep}) becomes identical to the corresponding
$\epsilon\chi$PT partition function \cite{ADOS}
\be
{\cal Z}_{\chi PT}
={\int dU(N_f) \det[U]^\nu}
\exp\left[{\Tr \frac{1}{4}{\mu^2 F_\pi^2V}{[U,B][U^\dagger, B]}
+ \frac12V \Sigma{M_f}( {U}+ {U^\dagger})}\right]\ .
\label{ZechPT}
\ee
$F_\pi$ and $\Sigma$ 
have as source terms chemical potential
through the charge matrix 
$B=$diag$(\mu_1\one_{N_1},\mu_2\one_{N_2})$,
and the diagonal mass matrix 
$M_f=$diag$(\{m_{f1}\},\{m_{f2}\})$, respectively.
For explicit results for these partition functions we
refer to \cite{ADOS}.

\section{Results for individual Dirac eigenvalue distributions}
\label{results}

In the following we first define all density correlations, all individual
eigenvalues correlations (or gap
probabilities), and then express the latter in terms of the former.
This inversion relation is valid for any theory expressed in
terms of Dirac eigenvalues, having a jpdf ${\cal P}(\{x\},\{y\})$
that is symmetric under exchange of
all $x_i$ and all 
$y_i$ eigenvalues separately. This applies to our chR2MT 
eq. (\ref{evrep}), its equivalent $\epsilon\chi$PT, 
or a Lattice QCD partition functions in terms of Dirac
eigenvalues.   
   
All density correlation functions are defined by integrating
all but $k(l)$ eigenvalues of ${\cal D}_1 ({\cal D}_2)$
\bea
R_{k,l}(x_1,\ldots,x_{k},\ y_1,\ldots,y_{l})&\equiv&
\frac{N!^2}{(N-k)!(N-l)!{\cal Z}}
\int_0^{\infty}  \prod_{i=k+1}^N dx_i \prod_{j=l+1}^N dy_j 
{\cal P}(\{x\},\{y\})\ .
\label{Revrep}
\eea
The simplest nontrivial example is the probability density $R_{1,1}(x,y)$ 
for finding an eigenvalue of
${\cal D}_1$  at $x$ and of ${\cal D}_2$  at $y$. 
If all eigenvalues of one kind are
integrated out one finds back the known quantities of the one-matrix theory
at $\mu=0$ \cite{ADOS}.  
Next we define the following gap probabilities that
the interval 
$[0,s]$ is occupied by $k$ eigenvalues 
and $[s,\infty)$ 
by $(N-k)$ eigenvalues of ${\cal D}_1$, and that the interval 
$[0,t]$ is occupied by $l$ eigenvalues 
and $[t,\infty)$ 
by $(N-l)$ eigenvalues of ${\cal D}_2$:
\bea
E_{k,l}(s,t) &\equiv& \frac{N!^2}{(N-k)!(N-l!){\cal Z}}
\int_0^s dx_1\ldots dx_{k}\int_s^\infty dx_{k+1}\ldots dx_N 
\int_0^t dx_1\ldots dx_{l}\int_t^\infty dy_{l+1}\ldots dy_N 
\nn\\
&&\times\ {\cal P}(\{x\},\{y\})
\ , \ \ \mbox{for}\ k,l=0,1,\ldots,N\ \ .
\label{Ekl}
\eea
The simplest example is $E_{0,0}(s,t)$ to find the intervals $[0,s]$ and 
$[0,t]$ empty of ${\cal D}_1$- and ${\cal D}_2$-eigenvalues, respectively.
Similarly we can define the probability to find the $k$-th 
${\cal D}_1$-eigenvalue 
at value $x_k=s$, and the $l$-th ${\cal D}_2$-eigenvalue 
at value $y_l=t$, to be 
\bea
p_{k,l}(s,t) &\equiv& k {N \choose k}l {N \choose l}
\frac{1}{{\cal Z}}
\int_0^s dx_1\ldots dx_{k-1}\int_s^\infty dx_{k+1}\ldots dx_N 
\int_0^t dy_1\ldots dy_{l-1}\int_t^\infty dy_{l+1}\ldots dy_N 
\nn\\
&&\times\  
{\cal P}(x_1,\ldots,x_{k-1},x_k=s,x_{k+1},\ldots, x_N,
y_1,\ldots,y_{l-1},y_l=t,x_{l+1},\ldots, y_N)\ ,
\label{pkl}
\eea
where the eigenvalues are ordered $x_1\leq\ldots\leq x_N$ and
$y_1\leq\ldots\leq y_N$. The simplest example is $p_{1,1}(s,t)$, the
distribution of each first eigenvalue. 
It is easy to see \cite{AD07} that all the quantities eq. (\ref{pkl}) can be
obtained from eq. (\ref{Ekl}) by taking two derivatives, 
\be
 \frac{\partial^2}{\partial s\partial_t} E_{k,l}(s,t)
\ =\ k!\ l!\left( p_{k,l}(s,t) - p_{k+1,l}(s,t)
-p_{k+1,l}(s,t) + p_{k+1,l+1}(s,t)\right)\ .
\label{Eklpkl}
\ee
Here, we define $p_{k,l}=0$ whenever index $k$ or $l$ is zero.
Finally we give an inversion formula expressing all gap probabilities,
and hence all individual eigenvalue distributions in terms of densities:
\be
E_{k,l}(s,t)= \sum_{i=0}^{N-k}\sum_{j=0}^{N-l} \frac{(-)^{i+j}}{i!j!} 
\int_0^s  dx_1\ldots dx_{k+i}\int_0^t  dy_1\ldots dy_{l+j}
R_{(k+i,l+j)}(x_1,\ldots,x_{k+i},\ y_1,\ldots,y_{l+j}).
\label{EklR}
\ee
The derivation \cite{AD07} follows closely the $\mu=0$ case \cite{ADp}. Since
it is known how to generate all higher density correlations $R_{k,l}$
from resolvents by inserting additional, auxiliary pairs of fermions and
bosons e.g. into eq. (\ref{ZechPT}), this relation clarifies how to generate
individual eigenvalue distributions in this setting from field theory.

\section{Examples
}\label{ex}

\begin{figure}[-h]
\includegraphics[width=.49\textwidth]{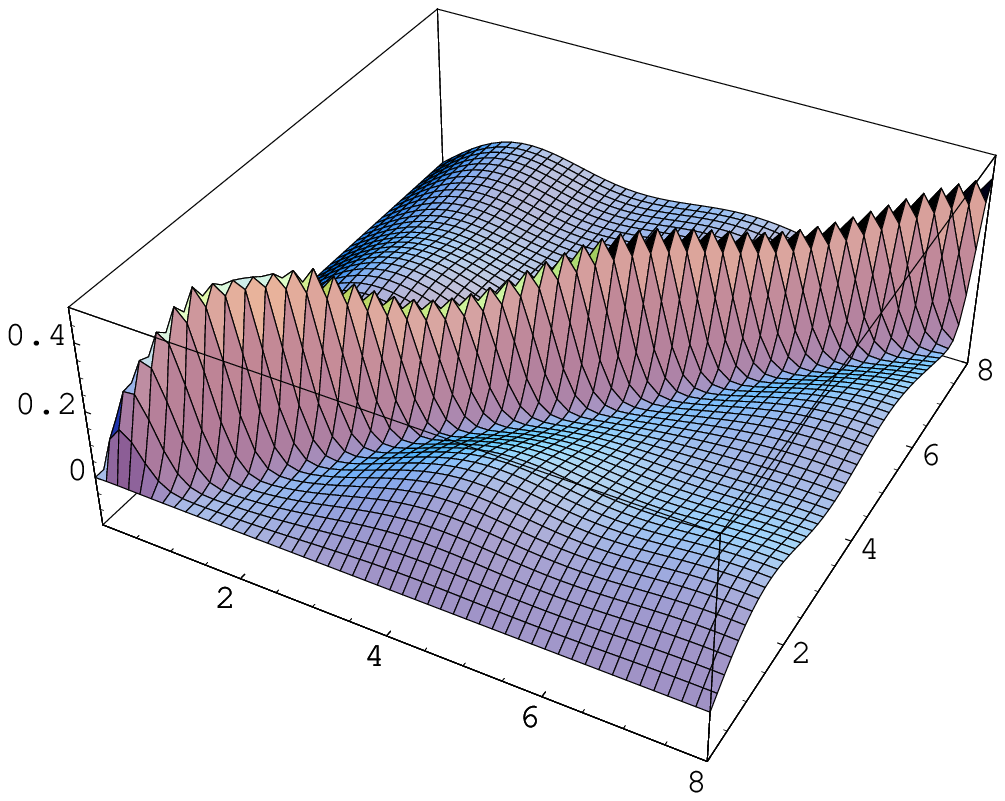}
\includegraphics[width=.49\textwidth]{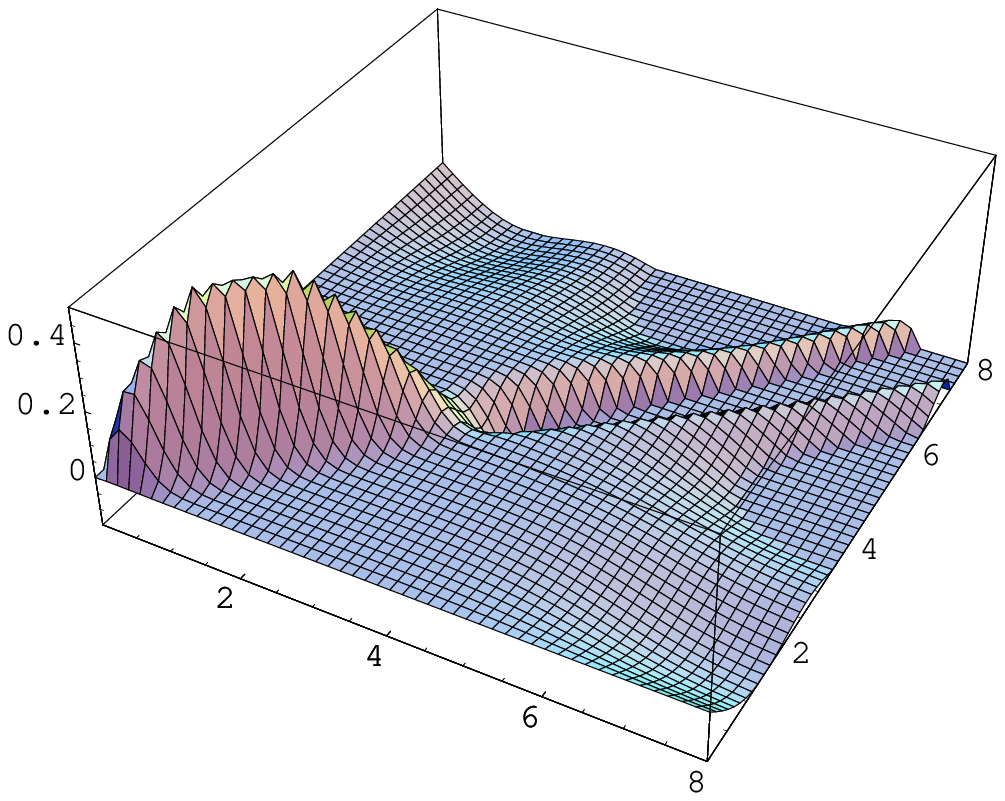}
\includegraphics[width=.49\textwidth]{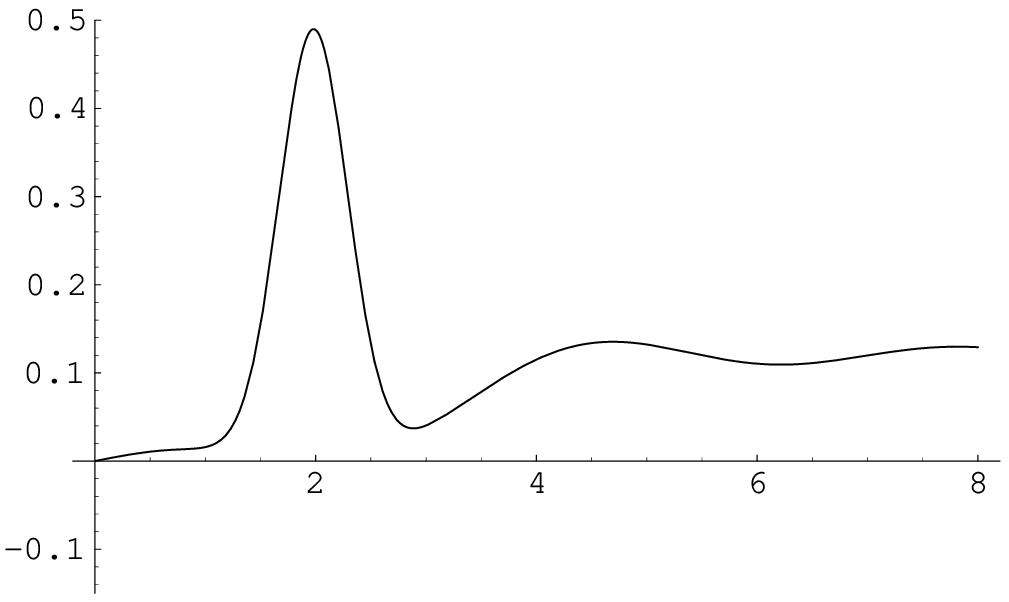}
\includegraphics[width=.49\textwidth]{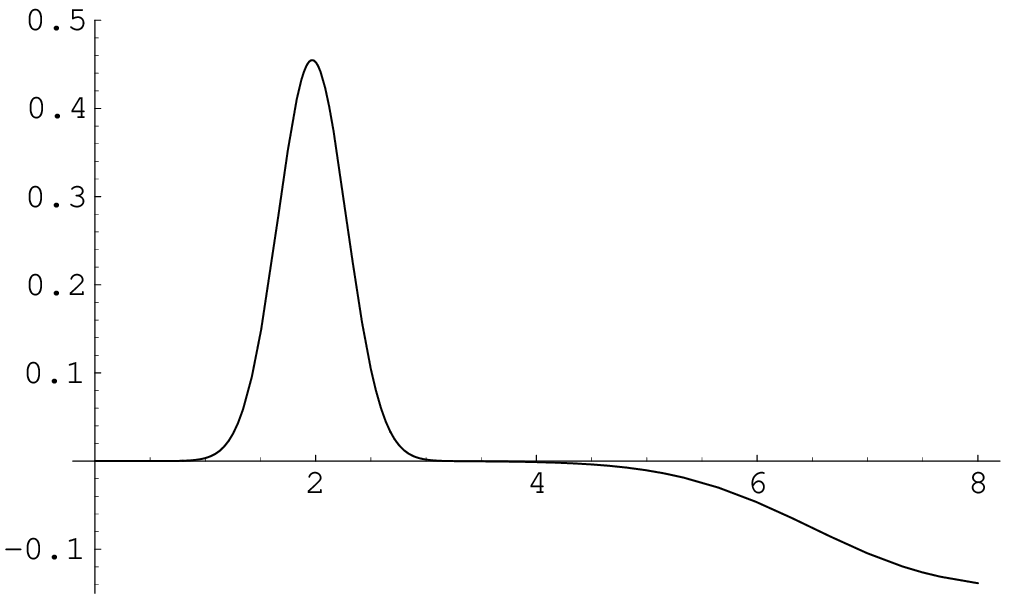}
\caption{Example quenched
density $R_{1,1}(s,t)$ (top left) vs individual eigenvalue distribution
$p_{1,1}(s,t)$ (top right)
at $\nu=0$ and $2\hmu=0.159$. 
The lower plots show corresponding 2D cuts at fixed $s=2$. 
}
\label{R11quenchnu0}
\end{figure}

\begin{figure}[-h]
\includegraphics[width=.49\textwidth]{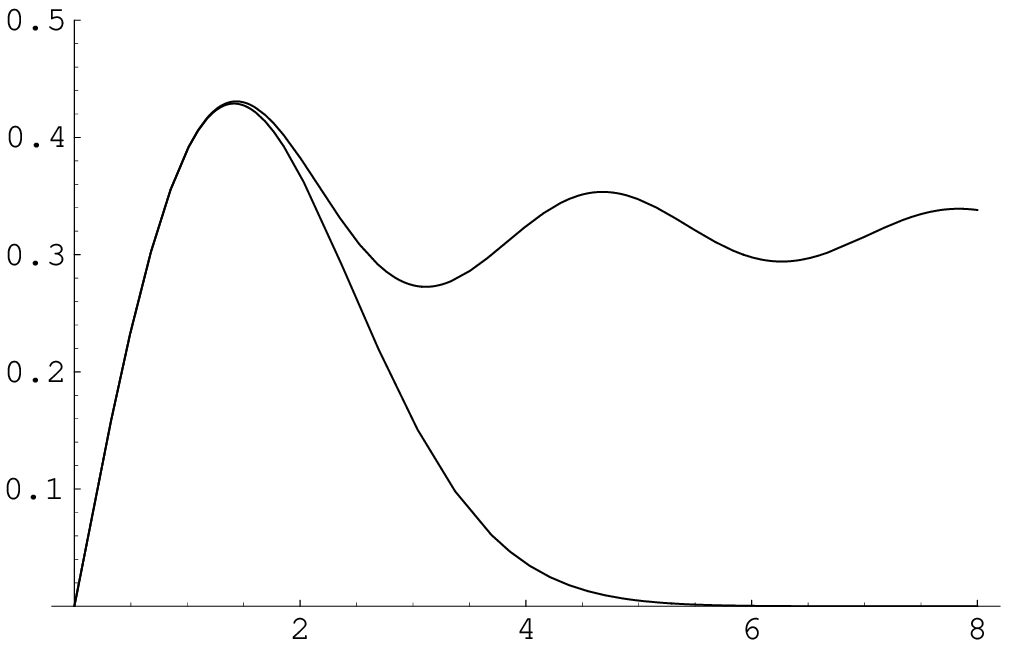}
\put(-120,120){$\rho_{1,0}(\hx)$}
\put(-170,45){$p_{1}(\hx)$}
\includegraphics[width=.49\textwidth]{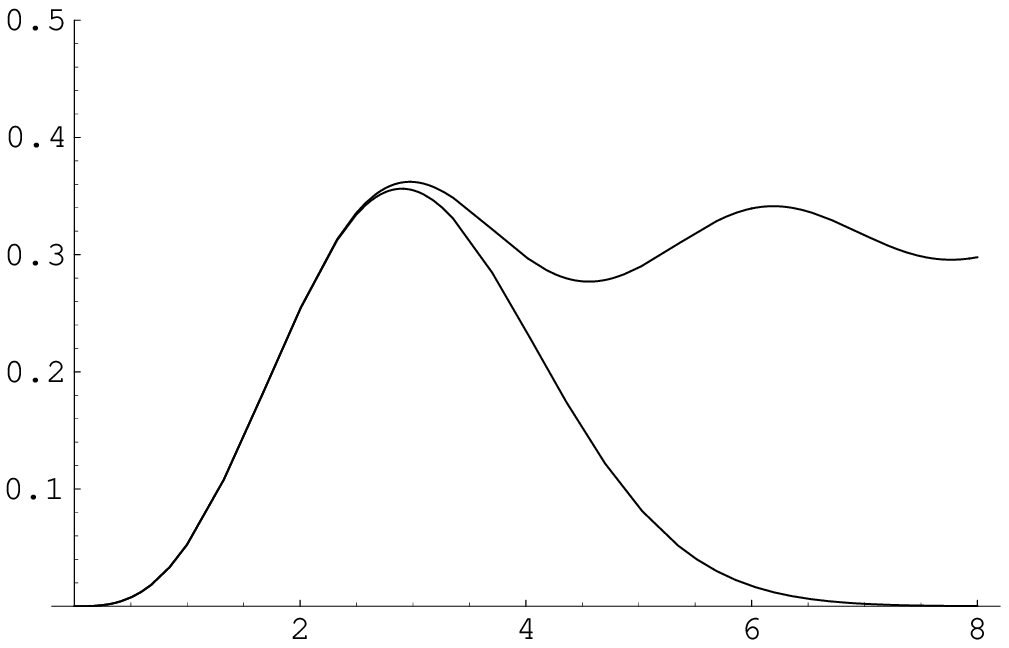}
\put(-80,120){$\rho_{1,0}(\hx)$}
\put(-140,45){$p_{1}(\hx)$}
\caption{
The one-matrix density $\rho_{(1,0)}(\hx)$
vs. the exact distribution of the first eigenvalue for $\nu=0$ :
$p_1(\hx)=\frac12\hx\ e^{-\frac14 \hx^2}$ (left),  and for $\nu=1$ : 
$p_1(\hx)=\frac12\hx\ e^{-\frac14 \hx^2}I_2(\hx)$
(right).}
\label{rhoreal}
\end{figure}

\begin{figure}[-h]
\includegraphics[width=.49\textwidth]{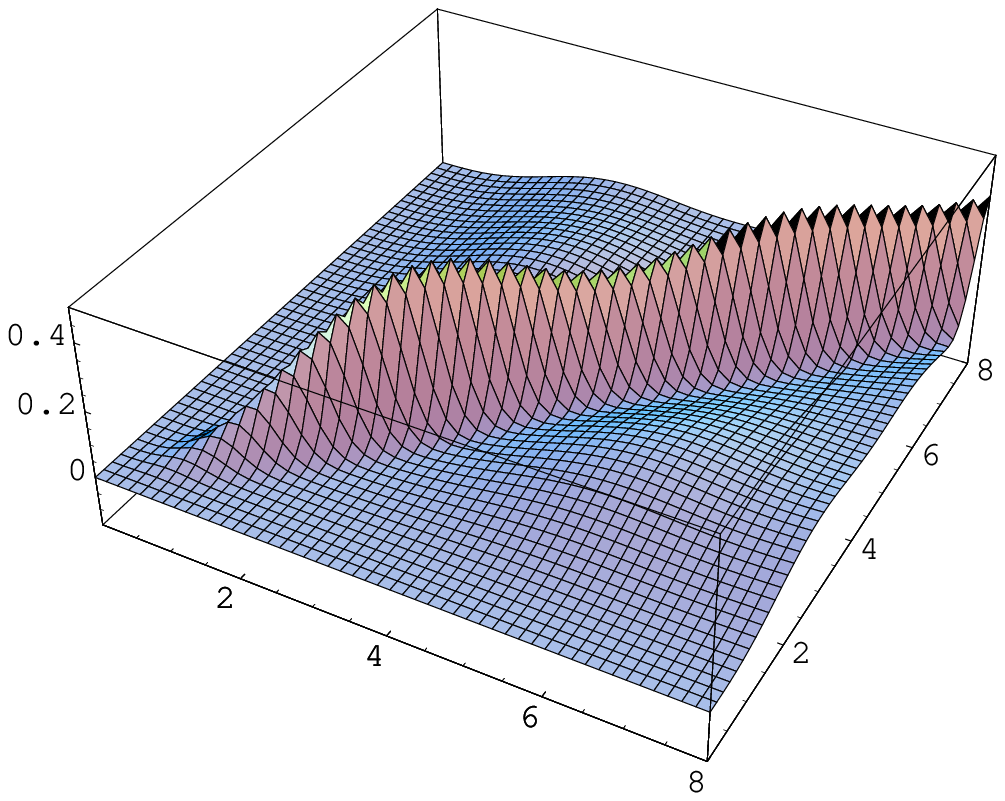}
\includegraphics[width=.49\textwidth]{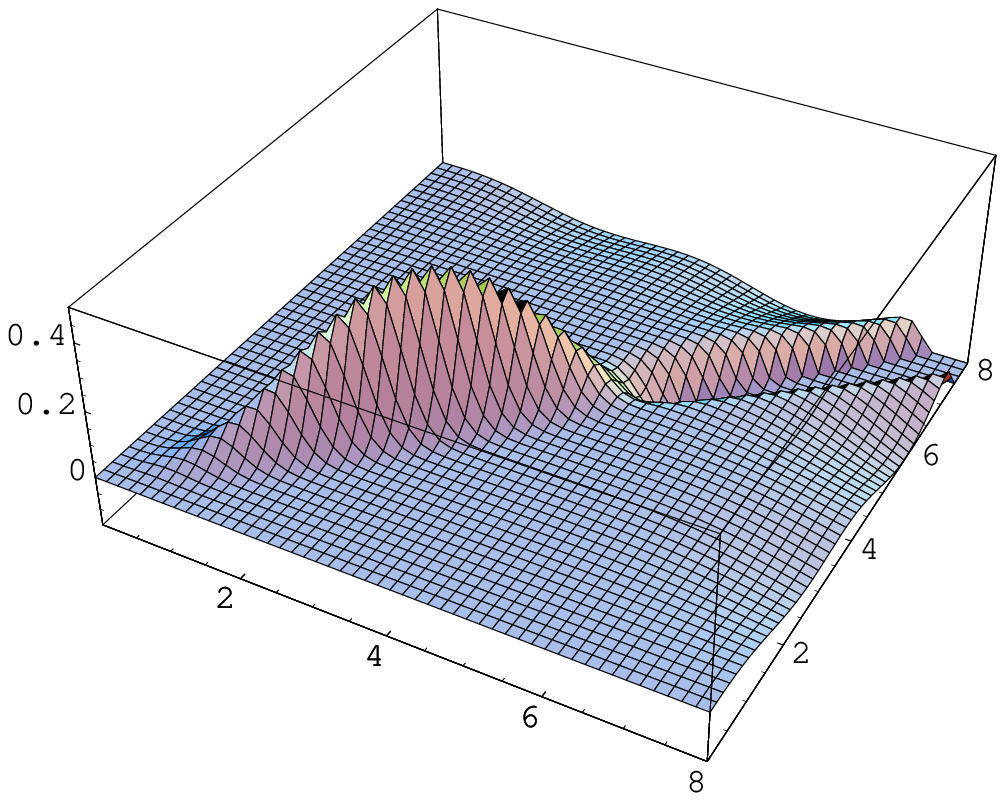}
\includegraphics[width=.49\textwidth]{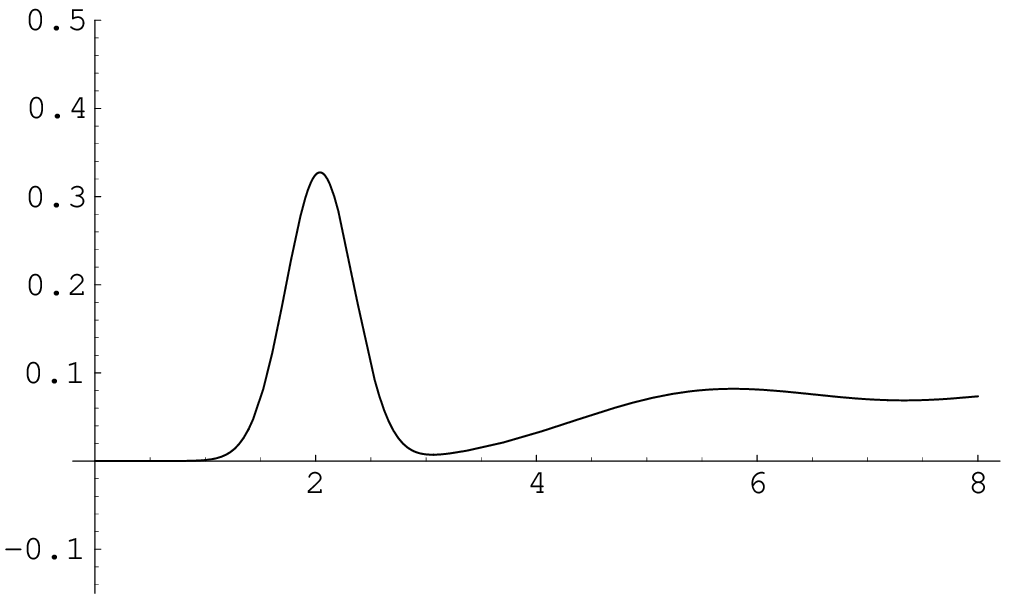}
\includegraphics[width=.49\textwidth]{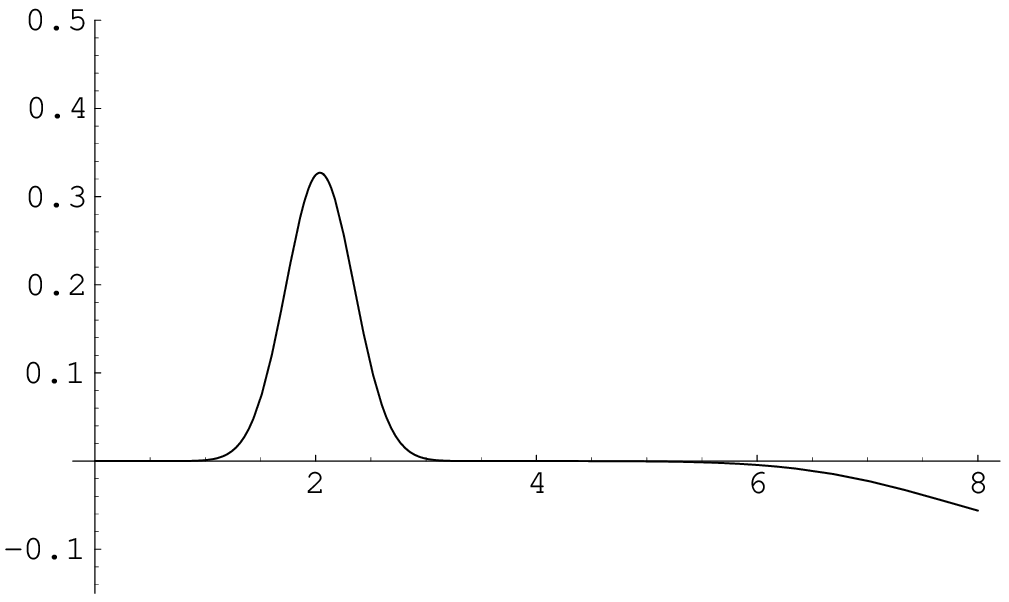}
\caption{Same as fig. 1. for $\nu=1$. 
The exact zero eigenvalues push the density away from the origin. 
}
\label{R11quenchnu1}
\end{figure}

We discuss in detail the simplest example, the probability $p_{1,1}(s,t)$. 
It follows from
the gap probability using eq. (\ref{Eklpkl}):
\be
\frac{\partial^2}{\partial s\partial t} E_{0,0}(s,t)\ =\ p_{1,1}(s,t)\ .
\label{Ep00}
\ee
We expand $E_{0,0}(s,t)$ to include at most 3-point density correlations as an
approximation, 
\bea
E_{0,0}(s,t) &=&
1-\int_0^s dx\, R_{1,0}(x)- \int_0^t dy\, R_{0,1}(y)
+ \int_0^s dx\int_0^t dy  R_{1,1}(x,y)\nn\\
&&+ \frac12 \int_0^t dy_1dy_2\, R_{0,2}(y_1,y_2)
+ \frac12 \int_0^s dx_1dx_2\, R_{2,0}(x_1,x_2)\nn\\
&&- \frac12 \int_0^s dx_1dx_2 \int_0^t dy\, R_{2,1}(x_1,x_2,y)
- \frac12 \int_0^s dx\int_0^t dy_1dy_2\, R_{1,2}(x,y_1,y_2) \ +\ \ldots
\label{E00ex} 
\eea
The derivatives eliminate all integrals over one-matrix densities 
that only depend on $s$ or $t$: 
\be
p_{1,1}(s,t)\ =\ R_{1,1}(s,t)\ -\  \int_0^s dx\, R_{2,1}(x,s,t)
\ -\ \int_0^t dy\, R_{1,2}(s,t,y)\ +\ \ldots\ .
\label{p00ex}
\ee
The leading order term is obviously given by the density 
$R_{1,1}(s,t)$, as can be clearly seen in figs.  
\ref{R11quenchnu0} and \ref{R11quenchnu1}.
There, we display the quenched density in the case of imaginary
isospin chemical potential $\mu_1=-\mu_2\equiv-\mu$. In the
microscopic large-$N$ limit $\rho_{1,1}(\hx,\hy)=\lim_{N\to\infty}R_{1,1}
(x=\hx/N,y=\hy/N)$ we obtain the following result \cite{DHSST,ADOS}
\bea
\rho_{(1,1)}(\hat{x},\hat{y})&=&
\rho_{(1,0)}(\hx)\rho_{(0,1)}(\hy)-
\hat{x}\hat{y}{\cal K}^+(\hy,\hx)\left({\cal
  K}^-(\hx,\hy)-\frac{1}{4\hmu^2}I_\nu\left(\frac{\hx\hy}{4\hmu^2}\right) 
e^{-\frac{\hx^2+\hy^2}{8\hmu^2}}\right)\nn\\
{\cal K}^\pm(\hx,\hy) &\equiv& 
 \int_0^1 dtt \, e^{\pm2\hmu^2
   t^2}J_{\nu}(\hx t)J_{\nu}(\hy t)\ ,\ \ 
{\cal K}^0(\hx,\hy) \equiv 
 \int_0^1 dtt \, J_{\nu}(\hx t)J_{\nu}(\hy t)\ .
\label{rhoQ11}
\eea
Here also the well known one-matrix density appears,
\be
\rho_{(1,0)}(\hat{x}) =\rho_{(0,1)}(\hat{x})
=\frac{\hat{x}}{2}\left[J_{\nu}^2(\hat{x})
- J_{\nu+1}(\hat{x})J_{\nu-1}(\hat{x})\right] 
\ =\ {\cal K}^0(\hx,\hx)~, \label{rhoQ}
\ee
see fig. \ref{rhoreal}.
Eq. (\ref{rhoQ11}) was derived independently for the chR2MT
eq. (\ref{evrep}) 
\cite{ADOS} and prior to that for $\epsilon\chi$PT 
eq. (\ref{ZechPT}) using replicas and the Toda-lattice
hierarchy \cite{DHSST}. It is displayed in figs. 
\ref{R11quenchnu0} and \ref{R11quenchnu1} left
for topological charge $\nu=0$ and 1, respectively, 
including 2-dimensional cuts. Because the density is the expectation value
$R_{1,1}(x,y) \sim \left\langle\  \Tr 
\delta({\cal D}_1-x)\ \Tr \delta({\cal D}_2-y)\ \right\rangle$,
$\mu\neq0$ resolves the delta function $\delta(x-y)$ that we would obtain at
$\mu=0$, times the one-matrix density eq. (\ref{rhoQ}) that we give for
comparison.  

Next we move to individual eigenvalues. 
A closed determinantal 
expression for all higher density correlation functions in terms of
the same building blocks as in eq. (\ref{rhoQ11}) was given in \cite{ADOS}:
\be
\rho_{k,l}(\{{\hx}\},\{{\hy}\}) =
\prod_{i}^k x_i \prod_{j}^l y_j
\det
\!\left[
\begin{array}{cc}
{\mathcal{K}}^0(\hx_{i_1},\hx_{i_2})
&
\ \ {\mathcal{K}}^-(\hx_{i_1},\hy_{j_2})-\frac{1}{4\hmu^2}
I_\nu\left(\frac{\hx_{i_1}\hy_{j_2}}{4\hmu^2}\right) 
e^{-\frac{\hx_{i_1}^2+\hy_{j_2}^2}{8\hmu^2}}
\\
{\mathcal{K}}^+(\hy_{j_1},\hx_{i_2})
&
{\mathcal{K}}^0(\hy_{j_1},\hx_{j_2})
\end{array}
\right].
\ee
We can insert these formulas into the expansion
eq. (\ref{p00ex}), after taking the microscopic limit. The result truncated at
the given order
is plotted in figs. 
\ref{R11quenchnu0} and \ref{R11quenchnu1} right. 
The fact that the truncated sum is an approximation is seen from the fact that
the individual eigenvalue density becomes negative (or diverges when adding
higher order terms). 
For the given values in the figs. this happens above $s=t\approx4$, and 
we have cut the 3D plots at values below $-0.15$.
Higher order terms in the
expansion eq. (\ref{EklR}) will keep 
the individual eigenvalue distribution to be zero for larger values of $s$ and
$t$. 
From our experience with the case $\mu=0$ \cite{ADp} we expect that this
expansion converges fast. 
The next to 
leading order used in the figures gives already a reasonably good
approximation.

\section{Conclusions and outlook}\label{conc}

We have shown how to derive individual eigenvalues distributions 
$p_{k,l}$ for two sets
of real Dirac operators with different imaginary chemical
potentials $\mu_{1,2}$. 
Similar expressions have been derived for a single Dirac operator
with complex eigenvalues at real $\mu$ and compared to Lattice data
\cite{ABSW}.
Both types of $\mu$ couple to $F_\pi$ and thus allow to fully determine all
LECs in the leading order $\chi$PT Lagrangian.
But only imaginary $\mu$ with real Dirac 
eigenvalues allow to date to perform unquenched or partially
quenched simulations.
The equivalence of $\epsilon\chi$PT to the chR2MT we mentioned here for the
density and partition function has been derived
very recently for all correlators \cite{BA}.

We have given an effective expansion for the distributions $p_{kl}$ by
truncating the sum over integrated densities, as was illustrated in our
examples. The possibility to derive exact expressions (which is possible for
real $\mu$) is currently under investigation \cite{AD07}.  
Our hope is that the results presented here will become as useful as
previously for $\mu=0$.

\acknowledgments
This work was supported by 
EPSRC grant EP/D031613/1 (G.A.) and 
EU network ENRAGE MRTN-CT-2004-005616.

\end{document}